\PassOptionsToPackage{dvipsnames,svgnames,x11names}{xcolor} 

\documentclass[sigplan,screen]{acmart}

\usepackage{amsmath}
\usepackage{subcaption}
\usepackage{supertabular}
\usepackage{stmaryrd}
\usepackage{xcolor}
\usepackage{multirow}
\usepackage{calc}
\usepackage[T1]{fontenc}
\usepackage{graphicx}
\usepackage{xspace}
\usepackage{dashrule}  
\usepackage{stackrel}
\usepackage{enumerate}
\usepackage{framed}
\usepackage{bussproofs}
\usepackage{mdframed}  
\usepackage{comment}
\usepackage{ifthen}
\usepackage{pdftexcmds}
\usepackage{mathpartir}
\usepackage{microtype}
\usepackage{fancyvrb}
\usepackage{booktabs}
\usepackage{tikz}
\usepackage{tikz-cd}
\usepackage[zi4]{inconsolata-slanted} 
\usepackage{listings}
\usepackage{url}
\usepackage{float}
\usepackage{subfiles}
\usepackage{csquotes}
\usepackage{balance}

\usetikzlibrary{calc}
\tikzset{curve/.style={settings={#1},to path={(\tikztostart)
    .. controls ($(\tikztostart)!\pv{pos}!(\tikztotarget)!\pv{height}!270:(\tikztotarget)$)
    and ($(\tikztostart)!1-\pv{pos}!(\tikztotarget)!\pv{height}!270:(\tikztotarget)$)
    .. (\tikztotarget)\tikztonodes}},
    settings/.code={\tikzset{quiver/.cd,#1}
        \def\pv##1{\pgfkeysvalueof{/tikz/quiver/##1}}},
    quiver/.cd,pos/.initial=0.35,height/.initial=0}

\usepackage{lstparams}
\usepackage{lstcoq}

\newcommand\informed{informed }

\usepackage[capitalize, nameinlink]{cleveref}

\usepackage{siunitx}
\newcommand\cdh[1]{\lstinline[language=Haskell,breakatwhitespace]{#1}}

\newcommand\cdc[1]{\lstinline[language=Coq,breakatwhitespace]{#1}}
\newcommand\zcdc[1]{\let\par\endgraf\cdc{#1}}

\newcommand\hstocoq{\texttt{hs-to-coq}\xspace}

\newcommand\etc{\textit{etc.}}
\newcommand\ie{\textit{i.e.,\ }}
\newcommand\eg{\textit{e.g.,\ }}

\newboolean{extended}
\newboolean{review}

\usepackage{draft} 
\newnote{scw}{blue}       
\newnote{ly}{blue}      
\newnote{lx}{olive}      
\newnote{bb}{ForestGreen}      
\newnote{sx}{pink} 

\draftfalse
\extendedfalse
\reviewfalse

\setcopyright{cc}
\setcctype{by}

\acmDOI{10.1145/3759425.3763391}
\acmYear{2025}
\copyrightyear{2025}
\acmISBN{979-8-4007-2148-9/25/10}
\acmConference[LMPL '25]{Proceedings of the 1st ACM SIGPLAN International Workshop on Language Models and Programming Languages}{October 12--18, 2025}{Singapore, Singapore}
\acmBooktitle{Proceedings of the 1st ACM SIGPLAN International Workshop on Language Models and Programming Languages (LMPL '25), October 12--18, 2025, Singapore, Singapore}
\received{2025-07-07}
\received[accepted]{2025-08-08}

\begin{document}
\title{A Case Study on the Effectiveness of LLMs in Verification with Proof Assistants}

\author{Bar{\i}\c{s} Bayaz{\i}t}
\email{baris@cs.toronto.edu}
\orcid{0009-0001-1094-1563}
\affiliation{%
  \institution{University of Toronto}
  \city{Toronto}
  \state{Ontario}
  \country{Canada}
}

\author{Yao Li}
\email{liyao@pdx.edu}
\orcid{0000-0001-8720-883X}
\affiliation{%
  \institution{Portland State University}
  \city{Portland}
  \state{Oregon}
  \country{USA}
}

\author{Xujie Si}
\email{six@cs.toronto.edu}
\orcid{0000-0002-3739-2269}
\affiliation{%
  \institution{University of Toronto}
  \city{Toronto}
  \state{Ontario}
  \country{Canada}
}

\begin{abstract}
  Large language models (LLMs) can potentially help with verification using
  proof assistants by automating proofs. However, it is unclear how effective
  LLMs are in this task. In this paper, we perform a case study based on two
  mature Rocq projects: the \hstocoq tool and Verdi. We evaluate the
  effectiveness of LLMs in generating proofs by both quantitative and
  qualitative analysis. Our study finds that: (1)~external dependencies and
  context in the same source file can significantly help proof generation;
  (2)~LLMs perform great on small proofs but can also generate large proofs;
  (3)~LLMs perform differently on different verification projects; and (4)~LLMs
  can generate concise and smart proofs, apply classical techniques to new
  definitions, but can also make odd mistakes.
\end{abstract}

\ccsdesc[500]{Software and its engineering~Formal software verification}
\ccsdesc[500]{Theory of computation~Automated reasoning}
\ccsdesc[300]{Theory of computation~Program verification}

\keywords{proof assistants, Coq (Rocq), large language models, program verification, interactive theorem proving, proof automation}
\maketitle

\bibliographystyle{ACM-Reference-Format}

\section{Introduction}\label{sec:intro}

Software should be correct. But in reality, that's rarely true.

Proof assistants allow us to formally \emph{verify} that a class of bugs is
\emph{absent} in a program via mechanized mathematical proofs. In the past two
decades, various works have demonstrated that this approach is a feasible way to
ensure software correctness and reliability. Some notable verified software
includes the CompCert C compiler~\citep{compcert}, the seL4
microkernel~\citep{sel4}, the CertiKOS operating system~\citep{certikos}, the
FSCQ file system~\citep{fscq},~\etc\@ Many new tools and frameworks that support
mechanized reasoning have also emerged, including program logics and frameworks
for reasoning about concurrency~\citep{iris,iris4,csl-sem,csl}, nonterminating
programs~\citep{itree,dijkstra-forever},
nondeterminism~\citep{omni-semantics,incorrectness,rev-hoare}, \etc\@

However, despite all these efforts, proving the correctness theorems of a program in
a proof assistant remains a daunting task. For example, \citet{containers}'s
work on verifying Haskell's \cdh{containers} library using \hstocoq shows that
their verification work ``required 8.9 lines of proof per line of
code''~\citep[Section 3]{containers}. This is a significant overhead in addition
to code development. Future changes to the code or the specification bring even
greater challenges for proof maintenance and proof
repair~\citep{sisyphus,qed-at-large,proof-repair}.

Large Language Models, or LLMs, on the other hand, have received great attention
for their capability in performing a wide range of tasks. In particular, existing
works have demonstrated LLMs' effectiveness in generating
code~\citep{codegen-llm-survey} and mathematical
proofs~\citep{AlphaGeometry,AlphaProof,OlympiadIneq}.

It is natural to ask: Can LLMs help with verification using proof assistants?

Indeed, researchers have recently started investigating this question, and
various new tools/frameworks for generating program correctness proofs with the
help of LLMs have also emerged~\citep{palm, fscq-llm, baldur, coq-pilot}.
However, due to the mysterious nature of LLMs~\citep{human-centric,
  llm-explain-survey}, many questions remain unanswered.

In this paper, we build on prior work and try to understand more about the
effectiveness of LLMs in verification with proof assistants, by conducting a
case study\footnote{The source code is publicly available at \href{https://github.com/bbayazit16/lmpl-2025-artifact}{https://github.com/bbayazit16/lmpl-2025-artifact}.} on two different verification projects that use Rocq
Prover~\citep{coq}: the \hstocoq tool~\citep{hs-to-coq, containers} and
Verdi~\citep{verdi, verdi-raft}.

Our case study investigates the following research questions:
\begin{itemize}
  \item \textbf{RQ1:} How do external dependencies and/or context in the same
        source file impact proof generation for a theorem?
  \item \textbf{RQ2:} How do LLMs perform on proofs of different sizes?
  \item \textbf{RQ3:} Is there a difference when running LLMs on different
        verification projects?
  \item \textbf{RQ4:} How is the quality of proofs generated by LLMs?
\end{itemize}

To answer these questions, we conduct a \emph{quantitative} study for RQ1, RQ2,
and RQ3, and a \emph{qualitative} study for RQ4. Our case study shows:
\begin{itemize}
  \item Including either external dependencies or context in the same source
        file, or both, can significantly improve the effectiveness of LLMs in
        generating proofs.
  \item LLMs perform significantly better on proofs of smaller sizes. However,
        there is still a chance for LLMs to generate proofs consisting of more than
        20 tactics.
  \item LLMs perform differently in the two projects we studied. For example,
        LLMs are less likely to generate proofs that are identical to original
        proofs in \texttt{hs-\allowbreak to-\allowbreak coq}. Context in the same source file also plays a more
        significant role in generating proofs for Verdi.
  \item LLMs can generate concise and smart proofs. They can also apply
        classical techniques such as performing a case analysis on an inductively
        defined proposition. On the other hand, LLMs can also
        generate odd, apparently failed proofs that repeat a tactic seemingly
        indefinitely.
\end{itemize}

In the rest of this paper, we first introduce some necessary background about
Rocq Prover and program verification in \cref{sec:background}. We then discuss
our target codebase, namely, \hstocoq and Verdi, in \cref{sec:codebase}. We
describe our methodology in \cref{sec:methodology}. We share and interpret our
results (both quantitative and qualitative ones) in \cref{sec:evaluation}.
We discuss related works in \cref{sec:related-work}. Finally, we conclude with
\cref{sec:conclusion}.

\section{Background}\label{sec:background}

\hyphenation{know-ledge}

In this section, we introduce the necessary background knowledge to understand this
paper. Readers who are familiar with these concepts should feel free to skip the
relevant parts.

\paragraph{Rocq Prover}
Some commonly used proof assistants include Rocq Prover~\citep{coq},
Agda~\citep{agda}, Lean~\citep{lean4}, F$\star$~\citep{fstar}, and
Isabelle~\citep{isabelle},~\etc\@ In this paper, we focus on Rocq Prover. Rocq
Prover is formerly known as the Coq proof assistant\footnote{The name change
  starts in Rocq Prover version 9.0. However, we will address all versions of
  Rocq Prover, including those before this name change, as Rocq Prover to
  avoid confusion.} and is one of the most commonly used proof assistants for
program verification. Rocq Prover has an expressive specification language and
supports full dependent types, which enables describing the properties of a software
system in rich detail.

We illustrate the process of theorem proving in Rocq Prover in
\cref{fig:coq-process}. We first state a theorem in Rocq Prover using its
specification language. For example, line~1 of \cref{fig:theorem-add_0_r} is
equivalent to the mathematical proposition:
$$\forall n \in \mathbb{N}, n + 0 = n$$
where $\mathbb{N}$ is the set of all natural numbers.

\begin{figure}[t]
\begin{subfigure}[b]{0.45\textwidth}
\begin{lstlisting}[language=coq,numbers=left,xleftmargin=4.0ex]
Theorem add_0_r : forall n:nat, n + 0 = n.
Proof.
  induction n as [ | n' IHn'].
  - (* n = 0 *)    reflexivity.
  - (* n = S n' *) simpl. rewrite -> IHn'.
    reflexivity.
Qed.
\end{lstlisting}
\caption{A Rocq theorem about natural numbers and its proof. The example comes
  from \emph{Logical Foundations}, a classical textbook on Rocq
  Prover~\cite{sf1}.}\label{fig:theorem-add_0_r}
\end{subfigure}

\begin{subfigure}[b]{0.45\textwidth}
\begin{lstlisting}[language=coq]
------------------------------
forall n : nat, n + 0 = n
\end{lstlisting}
\caption{The context and the proof goal when we enter the proof mode~(\ie~right
  after invoking line~2).}\label{fig:proof-mode}
\end{subfigure}

\begin{subfigure}[b]{0.45\textwidth}
\begin{lstlisting}[language=coq]
------------------------------
0 + 0 = 0
\end{lstlisting}
\caption{The context and the proof goal after we invoke the \cdc{induction}
  tactic~(\ie~after line~3). This is the first goal, \ie~the base case.}\label{fig:base-case}
\end{subfigure}

\begin{subfigure}[b]{0.45\textwidth}
\begin{lstlisting}[language=coq]
n' : nat
IHn' : n' + 0 = n'
------------------------------
S n' + 0 = S n'
\end{lstlisting}
\caption{The context and the proof goal after we proved the base case~(\ie~after
  line 4). This is the induction step.}\label{fig:induction-step}
\end{subfigure}

\begin{subfigure}[b]{0.45\textwidth}
\begin{lstlisting}[language=coq]
n' : nat
IHn' : n' + 0 = n'
------------------------------
S n' = S n'
\end{lstlisting}
\caption{The context and the proof goal after the \cdc{rewrite}
  tactic~(\ie~after line 5).}\label{fig:final-goal}
\end{subfigure}
\caption{The process of proving a theorem in Rocq Prover.}\label{fig:coq-process}
\end{figure}

We can then write a \emph{proof script} that instructs the proof assistant to
prove this theorem, as shown in lines~2--7. However, in Rocq Prover, we
typically do not directly write the entire proof script. Instead, we enter an
\emph{interactive} proof mode. This step is typically marked by the \cdc{Proof}
keyword~(line~2).

When we enter the proof mode, Rocq Prover displays the current context and
the proof goal as shown in \cref{fig:proof-mode}. The \emph{context}, which is empty
at this point, consists of all the current hypotheses. The \emph{proof goal} is what
we need to show to finish the proof. We can manipulate the context and the proof
goal using \emph{tactics}, which are instructions to Rocq Prover about how to
proceed with the proof. In this case, we decide to do an induction over $n$,
indicated by the tactic \cdh{induction n}~(line~3 in
\cref{fig:theorem-add_0_r}). Our tactic also names some new variables via an
\emph{intro pattern} \cdh{[ | n' IHn' ]}---these names will show up later in the
proof process.

After invoking the \cdc{induction} tactic, our proof goal will become two
subgoals: one for the base case and one for the induction step. Rocq Prover will
first ask us to prove the base case. We show the context and the goal of the
base case in \cref{fig:base-case}. We can see that the context is still empty,
but the goal has been changed to prove that $0 + 0 = 0$. Rocq Prover can tell
that $0 + 0$ computes to $0$, so proving the goal is equivalent to proving that
$0 = 0$. Rocq Prover knows that \cdc{=} is \emph{reflexive}, so we can discharge
this goal via the \cdc{reflexivity} tactic.

Once we are done with the base case, Rocq Prover will ask us to prove the
induction step. We show the context and the goal of the induction step in
\cref{fig:induction-step}. This time, the context contains a variable \cdc{n'}
that has type \cdc{nat} and an \emph{induction hypothesis} \cdc{IHn'} that
states \cdc{n' + 0 = n'}. The names \cdc{n'} and \cdc{IHn'} come from the intro
pattern in our \cdc{induction} tactic earlier. Our goal has also been changed to
show that \cdc{S n' + 0 = S n'}, where \cdc{S n'} means the successor of
\cdc{n'}.

In Rocq, \cdc{S n' + 0} is recursively defined as \cdc{S (n' + 0)}. We can
reveal this via the \cdc{simpl} tactic. After that, we recognize that \cdc{n' +
  0} equals \cdc{n'} by the induction hypothesis \cdc{IHn'}, so we can
\cdc{rewrite} using \cdc{IHn'}. The \cdc{rewrite} changes the goal to
\cref{fig:final-goal}, which can again be solved by the \cdc{reflexivity}
tactic.

Finally, we can write a \cdc{Qed} at the end of the proof~(line~7 in
\cref{fig:theorem-add_0_r}). \cdc{Qed} is more than an end mark of a proof in
Rocq Prover: it checks that the proof constructed by our proof script is indeed
a correct proof of the theorem. Rocq Prover has a small trusted computing base
for proof checking. This means that a proof checked by \cdc{Qed} is highly
trustworthy.

\paragraph{Program Verification with Rocq Prover}
We can verify the properties of a program in the same way in Rocq Prover. If the
program already exists but was written in another language, we need to first
\emph{embed} the syntax and/or semantics of that program in Rocq
Prover~\citep{embedding}. Alternatively, we can also write a program directly in
Rocq, prove properties about it, and then \emph{extract} it to another language
like OCaml or C~\citep{certicoq}. Our evaluation includes examples in both
approaches~(\cref{sec:codebase}).

There have been various verification works based on Rocq Prover, including
compilers~\citep{compcert, vellvm}, operating systems~\citep{certikos}, file
systems~\citep{fscq}, networked servers~\citep{itree-server, itree-kv-server},
cryptographic algorithm implementations~\citep{fiat-crypto},~\etc\@ There are
also various tools/frameworks supporting program verification with Rocq Prover,
including Verified Software Toolchains~\citep{vst}, Iris~\citep{iris, iris4},
certified abstraction layers~\citep{cal, ccal}, mathematical
components~\citep{mathcomp}, and interaction trees~\citep{itree,
  itree-layer},~\etc\@ A more detailed account of Rocq Prover's ecosystem can be
found in \citet{coq-ecosystem}.

\section{Codebase for Evaluation}\label{sec:codebase}

We chose two open-source Rocq Prover projects as the evaluation target: (1)~the
theory for Haskell's \cdh{base} library contained in the \hstocoq
project~\citep{hs-to-coq, containers, verify-ghc}, and (2)~the Verdi project for
implementing and verifying distributed systems\\~\citep{verdi, verdi-raft}.

We chose these two projects for the following reasons:

First, these two projects represent the two most typical ways to represent
programs: functions and inductively defined relations. Programs in the \hstocoq
codebase are all purely functional programs, so they are simply defined as Rocq
functions.\footnote{They should be called Gallina functions, to be more precise.
  Gallina is the specification language of Rocq Prover. However, we will not try
  to intentionally distinguish Gallina and Rocq in this paper.} The Verdi
project, on the other hand, reasons about traces and transition systems encoded
by inductively defined propositions.

Second, these two projects have proper sizes for an initial investigation of
LLMs' effectiveness in verification. On the one hand, they are no toy projects.
The theory of \cdh{base} in \hstocoq contains 187 proofs, and Verdi contains 579
proofs. On the other hand, these projects are not too large.

Finally, none of these projects involve advanced program logics~(\eg~separation
logic~\citep{separation-logic}, concurrent separation logic~\\\citep{csl,
  csl-sem}) or frameworks~(\eg~certified abstraction layers~\citep{cal, ccal},
interaction trees~\citep{itree, itree-layer}). The absence of these advanced
reasoning tools helps keep the experiments pristine.

We now discuss each project and the part we evaluate in more detail.

\subsection{The \hstocoq Project}

The \hstocoq tool translates purely functional programs in Haskell to a shallow
embedding in Rocq Prover. Its open-source repository contains translated code
from Haskell's \cdh{base} library, \cdh{containers} library, parts of the GHC
compiler, and many other examples of different sizes. We show an example of the
original Haskell code and the translated Rocq code in
\cref{fig:hs-to-coq-examples}. More details on how such a translation works can be
found in \citet{hs-to-coq}.

\begin{figure}[t]
\begin{subfigure}[b]{0.4\textwidth}
\begin{lstlisting}[language=haskell]
take                   :: Int -> [a] -> [a]
take n _      | n <= 0 =  []
take _ []              =  []
take n (x:xs)          =  x : take (n-1) xs
\end{lstlisting}
\caption{The \cdh{take} function defined in Haskell's \cdh{base} library~\citep{base-list-commit}.}
\end{subfigure}
\begin{subfigure}[b]{0.4\textwidth}
\begin{lstlisting}[language=coq]
Fixpoint take {a:Type} (n:Z) (xs:list a) :
  list a :=
  if (n <=? 0)%Z then nil
  else match xs with
      | nil => nil
      | cons y ys => cons y (take (n - 1) ys)
      end.
\end{lstlisting}
\caption{The \cdh{take} function converted to Rocq by \hstocoq. The
  \cdc{Fixpoint} keyword marks the definition of a recursive function. Haskell's
  types, such as \cdh{Int} and lists \cdh{[]}, are translated to Rocq types
  \cdc{Z} and \cdc{list}.}
\end{subfigure}
\caption{The \cdh{take} function defined in Haskell's \cdh{base} library and its
  translation in Rocq Prover produced by \hstocoq.}\label{fig:hs-to-coq-examples}
\end{figure}

Our case study is based on the translated Rocq code, and theorems stated and
proven by the \hstocoq developers. Once a piece of Haskell code is translated
using \hstocoq, we can treat the translated code as regular Rocq code, so our
evaluation does not rely on the \hstocoq tool or any Haskell code.

Our case study focuses on the theory of the \cdh{base} library. The \cdh{base}
library contains a number of basic Haskell types, functions, typeclasses, and
typeclass instances. The theory of \cdh{base} contains theorems for these basic
types and functions, and theorems for typeclass laws. 

Typeclasses are a way to implement \emph{overloading} (\ie~\emph{ad-hoc
  polymorphisms}) in functional languages, including both Haskell and Rocq
Prover~\citep{typeclass, typeclass-hs, typeclass-coq}. A few examples of
typeclasses implemented in the \cdh{base} library include: \cdh{Eq} for
equality tests, \cdh{Ord} for total orders, \cdh{Semigroup} for concatenation,
\cdh{Foldable} for ``congregating'' a data structure, and abstract interfaces
like \cdh{Functor}, \cdh{Applicative}~\citep{applicative}, and
\cdh{Monad}~\citep{moggi-monad, wadler-monad},~\etc\@

Instances of these typeclasses are expected to satisfy certain laws. For
example, an implementation of equality tests \cdh{==} in \cdh{Eq} should be
reflexive, transitive, and symmetric; the \cdh{<=} operator in \cdh{Ord} should
be reflexive, transitive, and antisymmetric; a \cdh{Monad} should satisfy monad
laws~\citep{moggi-monad, wadler-monad}. The documentation of the \cdh{base}
library describes these laws in details.

We show an example of how \hstocoq's theory of \cdh{base} states laws for the
\cdh{Eq} typeclass in \cref{fig:base-laws}. These laws are themselves defined as
typeclasses in Rocq Prover. \cdc{Eq_refl}, \cdc{Eq_sym}, and \cdc{Eq_trans}
state that \cdc{==} is reflexive, symmetric, and transitive, respectively. In
\hstocoq, \cdc{_==_} represents the equality test function, and \cdc{==} is a
notation that can be used as an infix operator. \cdc{Eq_inv} states that
\cdc{==} and \cdc{/=} are inverses of each other. Finally, \cdc{EqExact} contains
a special law that states \cdc{==} always agrees with Rocq's built-in equality
\cdc{=}. The \cdc{reflect} definition is an interesting definition that enables
a classical technique in mechanized reasoning called \emph{proof by reflection}.
We will see an example of LLMs using this later in \cref{sec:evaluation}.

\begin{figure}[t]
\begin{lstlisting}[language=coq]
Class EqLaws (t : Type) `{Eq_ t} :=
  { Eq_refl  : reflexive  _==_;
    Eq_sym   : symmetric  _==_;
    Eq_trans : transitive _==_;
    Eq_inv   : forall x y : t, x == y = ~~ (x /= y)
  }.

Class EqExact (t : Type) `{EqLaws t} :=
  { Eq_eq : forall x y : t,
      reflect (x = y) (x == y) }.
\end{lstlisting}
\caption{Laws for the \cdh{Eq} typeclass stated in Rocq Prover in
  \hstocoq~\citep{hs-to-coq, containers}.}\label{fig:base-laws}
\end{figure}

We choose the theory of \cdh{base} because it contains a fair amount of
theorems, and the proofs in general are neither too simple nor too complicated.
The longest proof script involves 43 tactics.

Other theories, such as theories for \cdh{containers}, \cdh{graph}, or the GHC
compiler, contain much more complicated proofs. For example, the theorem
\cdc{insertBM_Desc} is about the property of the \cdh{insertBM} function of
\cdh{container}'s \cdh{IntSet} data structure.\footnote{The data structure is a
  Patricia trie~\citep{mergeable-int-map, patricia}.} The handcrafted proof of
this theorem is 42 lines of proof script, makes heavy use of proven lemmas, uses
custom tactics, uses Ltac's \cdc{match} clause for pattern matching certain goals
to solve them automatically, involves both backward reasoning and forward
reasoning using \cdc{assert}. We leave the investigation of these examples to future
work.

The \hstocoq project relies on Rocq Prover 8.10, which is an old version first
released in April 2019. Unfortunately, most of its code no longer works under
later versions of Rocq Prover because Rocq Prover does not support backward
compatibility. For this reason, we conduct our study on Rocq Prover 8.10 as
well. However, the key workflow and features of Rocq Prover remain the same across
these versions.

\subsection{Verdi}\label{sec:verdi}

Verdi is a framework for implementing and verifying distributed systems in Rocq
Prover. Instead of writing a program in a different language and embedding it in
Rocq Prover, a programmer first implements their distributed systems in Rocq
Prover and extracts the code to OCaml. Unlike purely functional programs in
Haskell's \cdh{base} library, distributed systems always contain a number of
effects and interact with a network that can reorder or even drop messages. To
model this, Verdi defines a special monad for implementing distributed systems
and transition systems for network semantics. More details about how Verdi works
can be found in \citet{verdi, verdi-raft}.

The Verdi framework has been used in various works to study the effectiveness of
AI in verification. For example, it is included as part of the CoqGym
benchmark~\citep{coq-gym} and has been studied by \citet{diversity-driven,
  tactok}. In particular, \Citet{palm} tried applying
GPT-3.5\footnote{https://platform.openai.com/docs/models/gpt-3.5-turbo} to
proofs in Verdi. They found that LLMs like GPT-3.5 are ineffective in finishing
most of the proofs, as they collected 520 errors out of 579 theorems. They
further analyzed all the errors and made the following
observation~\citep[Section 3]{palm}:

\begin{displayquote}
  \ldots while LLMs often generate proof scripts with the right high-level
  structure, they often struggle with accurately addressing the sorts of
  low-level details that hammers excel at. For example, GPT-3.5 often knows when
  to use the induction tactic to decompose theorems into subgoals, but often
  fails to generate the right sequence of tactics to prove each subgoal\ldots
\end{displayquote}

This paper builds on these prior studies, but also investigates the
effectiveness of dependencies in prompting.

The Verdi project we experiment with is the version included in
CoqGym~\citep{coq-gym} and relies on Rocq Prover 8.11, to be consistent with
prior studies.

\section{Methodology}\label{sec:methodology}

To evaluate how models performed under different contexts, we extracted the
following information for each top-level construct using
SerAPI~\citep{GallegoArias2016SerAPI} version 8.10.0+0.7.2 for \hstocoq, and
8.11.0+0.11.1 for Verdi\footnote{Our experiments were conducted on Verdi
  corresponding to commit
  \lstinline|fdb4ede19d2150c254f0ebcfbed4fb9547a734b0|.}, along with Rocq
version 8.10.2 and 8.11.0, respectively:

\emph{Name and signature:} For each top-level definition in a Rocq source
file, we extracted its name (\ie~the identifier bound by the construct) and its
signature. For theorems, the signature consists of the entire declaration
excluding the proof. For other definitions, the signature includes the entire
definition.

\emph{In-file context:} We defined the in-file context as all lines in the
file prior to the location where a theorem appears.

\emph{External dependencies, or dependencies:} We defined external
dependencies~(or dependencies for short) as any \textit{signatures} that the
original proof relies on, including definitions and theorems from other source
files. If a dependency was already included in the in-file context, we excluded
it from the list of dependencies to avoid repetition.
Our extraction may include unnecessary dependencies. Specifically, qualified
identifiers returned by SerAPI can match identifiers defined in multiple files.
In such cases, we included all matching possibilities in the dependency
list.

\emph{Notations:} For each dependency and file imported via Rocq's \cdc{Require
  Import} command, we collected all associated notation declarations. However,
the definitions underlying these notations were not necessarily included as
dependencies, since a notation may be used without its underlying definition
being required by the proof.

\paragraph{Model and parameter selection}

Our model selection includes both general-purpose and reasoning models with a mix of full-sized and lightweight variants:
\begin{enumerate}
  \item GPT-4o-mini, version 2024-07-18: A smaller general-purpose model with a
        context length of \\ 128{,}000 tokens~\citep{OpenAI2024GPT4oMini}.
  \item GPT-4o, version 2024-11-20: A general-purpose model with a context
        length of 128,000 tokens~\citep{OpenAI2024GPT4o}.
  \item OpenAI o4-mini, version 2025-04-16: A smaller reasoning model with a
        context length of 200{,}000 tokens~\citep{OpenAIOpenAIO4Mini2025}. The
        model does not support changing the default temperature through the API,
        but supports a reasoning effort parameter
        \citep{azure_ai_foundry_reasoning_2025}. For our experiments, we have
        selected reasoning effort `medium,' which is the default.
  \item DeepSeek Prover V2: An open-source model based on DeepSeek V3. This
        model is fine-tuned for theorem proving in Lean 4. The model has a
        context length of 163{,}840 tokens~\citep{deepseek-prover-context} and a
        parameter count of 671 billion
        \citep{ren2025deepseekproverv2advancingformalmathematical}.
        We include this model in our case study to check if exposure to
        mechanized proofs in another proof assistant transfers to Rocq proofs.
  \item DeepSeek R1-0528: A large open-source reasoning model with a context length
        of 163{,}840 tokens~\citep{deepseek-r1-context}, and a parameter count
        of 671
        billion~\citep{deepseekai2025deepseekr1incentivizingreasoningcapability}.
\end{enumerate}

Each model was prompted with the same system message (for models supporting system prompt), and was allowed a maximum of 16{,}384 output tokens, configured using \lstinline|max_tokens| or \lstinline|max_completion_tokens| based on the model. The original context lengths for each model were preserved.

For all experiments, we set the temperature to 0.1 for models that support modifying this parameter over the API (e.g., GPT-4o). For models that do not support a custom temperature setting (e.g., o4-mini), the default value of 1.0 was used.

\paragraph{Prompt}

We used a minimal system prompt that described (1) the information provided to
the model, (2) the proof task it has to perform, and (3) the expected response
format, asking the model to respond only with the proof body. The prompt also
specified the current version of the Rocq available and included whether the
version used \lstinline[language=Coq]|omega| in place of
\lstinline[language=Coq]|lia|. We included this detail as the codebases being
evaluated were relatively old, whereas the models, which have more recent knowledge
cutoffs, are likely aware that \lstinline[language=Coq]|omega| is deprecated.

\paragraph{Variation of dependencies}\label{sec:variation}

We varied the prompt provided to the LLMs across four conditions: (1)~full
context (which we will shorten as \emph{the \informed mode} from now on),
(2)~without dependencies and notations, (3) without in-file context, and (4)
with both removed.

When omitting the in-file context, we still include the import statements
present in the file to show the model the available modules. We also extend
dependencies to include the in-file dependent signatures.

\paragraph{Checking successful proofs}
We defined a proof as successfully generated by the LLM if and only if SerAPI's
\lstinline|sertop| program accepted the proof when provided with (1) all lines
in the file preceding the theorem (\ie~\textit{the in-file context}), (2) the
theorem's signature, and (3) the LLM-generated proof body. This validation was
performed using the version of SerAPI that matches the Rocq Prover version used
in the corresponding codebase.

\section{Evaluation Results}\label{sec:evaluation}

We now share our evaluation results and use them to answer the four research
questions we proposed in \cref{sec:intro}.

\paragraph{RQ1: How do external dependencies and/or context in the same source
  file impact proof generation for a theorem?}

Among the four ablations we introduced in \cref{sec:variation}, most models
achieved the highest success rate in the \informed mode, as shown in
\cref{tab:basethy-rawdata,tab:verdi-rawdata}. For both \hstocoq and Verdi,
success rates dropped for most models when either in-file context or
dependencies were excluded, with the worst results occurring when both were
excluded.

One potential consequence of including all dependencies and in-file context is
an increase in input tokens. To understand this implication, we also estimated
the number of tokens required in both projects. We show the statistics in
\cref{tab:prompt-token-counts}.

\begin{table*}[ht]
    \centering
    \caption{Success counts and rates across different settings for \hstocoq and Verdi.}
    \label{tab:both-rawdata}
    \begin{subtable}[h]{\linewidth}
        \centering
        \caption{\hstocoq (187 theorems)}
        \label{tab:basethy-rawdata}
        \begin{tabular}{lcccc}
            \toprule
            Model           & Informed   & No in-file context & No dependencies & Neither \\
            \midrule
            GPT-4o-mini    & 42 (22.5\%)        & 35 (18.7\%)             & 44 (23.5\%)             & 31 (16.6\%)             \\
            GPT-4o          & 92 (49.2\%)        & 73 (39.0\%)             & 83 (44.4\%)             & 48 (25.7\%)             \\
            o4-mini  & 97 (51.9\%)        & 81 (43.3\%)             & 81 (43.3\%)             & 52 (27.8\%)             \\
            DeepSeek Prover V2
                            & 85 (45.5\%)        & 76 (40.6\%)             & 71 (38.0\%)             & 58 (31.0\%)             \\
            DeepSeek R1-0528     & 82 (43.9\%)        & 74 (39.6\%)             & 84 (44.9\%)             & 48 (25.7\%)             \\
            \bottomrule
        \end{tabular}
    \end{subtable}\vfill
    \begin{subtable}[h]{\linewidth}
        \centering
        \caption{Verdi (579 theorems)}
        \label{tab:verdi-rawdata}
        \begin{tabular}{lcccc}
        \toprule
        Model           & Informed  & No in-file context & No dependencies & Neither \\
        \midrule
        GPT-4o-mini     & 55 ( 9.5\%)        & 27 ( 4.7\%)             & 57 ( 9.8\%)             & 17 ( 2.9\%)             \\
        GPT-4o          & 177 (30.6\%)       & 117 (20.2\%)            & 170 (29.4\%)            & 38 ( 6.6\%)             \\
        o4-mini & 172 (29.7\%)       & 124 (21.4\%)            & 177 (30.6\%)            & 45 ( 7.8\%)             \\
        DeepSeek Prover V2
                        & 164 (28.3\%)       & 108 (18.7\%)            & 159 (27.5\%)            & 42 ( 7.3\%)             \\
        DeepSeek R1-0528     & 148 (25.6\%)       & 123 (21.2\%)            & 140 (24.2\%)            & 40 ( 6.9\%)             \\
        \bottomrule
    \end{tabular}
    \end{subtable}
\end{table*}

\begin{table}
  \centering
  \caption{Estimated prompt token counts for each setting, excluding the system prompt (rounded to the nearest integer). The token counts were estimated using OpenAI's TikToken library \citep{openai2025tiktoken}.}\label{tab:prompt-token-counts}
  \setlength{\tabcolsep}{4.03pt}
  \begin{tabular}{llrrr}
    \toprule
    \textbf{Project} & \textbf{Condition}         & \textbf{Mean} & \textbf{Median} & \textbf{Max} \\
    \midrule
    \multirow{4}{*}{\hstocoq}
      & Informed                      & 3379          & 3162   & 10223         \\
      & No dependencies           & 1766          & 1292      & 6833      \\
      & No in-file context                  & 1916          & 1862  & 5720           \\
      & Neither         & 152           & 147         & 228    \\
    \bottomrule
    \multirow{4}{*}{\texttt{Verdi}}
      & Informed                      & 6944          & 5488  &  25357         \\
      & No dependencies           & 5653          & 4393    & 19289        \\
      & No in-file context                  & 2559          & 1618  & 20674          \\
      & Neither         & 174           & 167   & 445          \\
    \midrule
  \end{tabular}
\end{table}

\paragraph{RQ2: How do LLMs perform on proofs of different sizes?}

\begin{figure}[t]
\begin{subfigure}[t]{0.45\textwidth}
    \centering
    \includegraphics[scale=0.53]{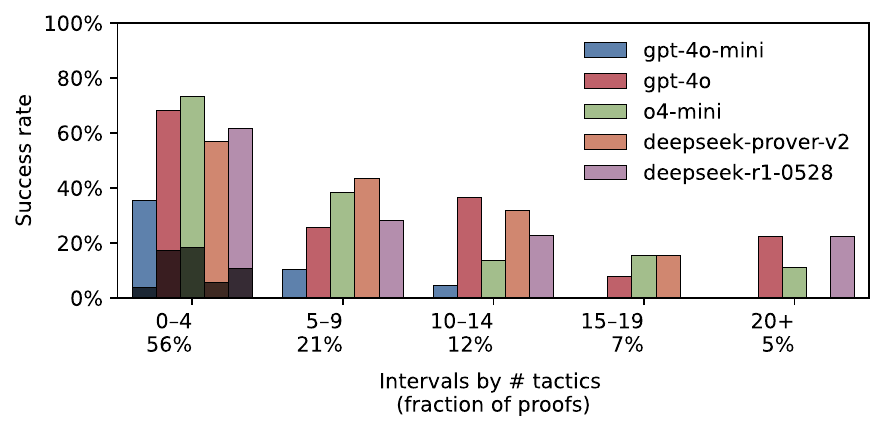}
    \caption{\hstocoq}\label{fig:basethy-samegraph}
  \end{subfigure}
  
\begin{subfigure}[t]{0.45\textwidth}
    \centering
    \includegraphics[scale=0.53]{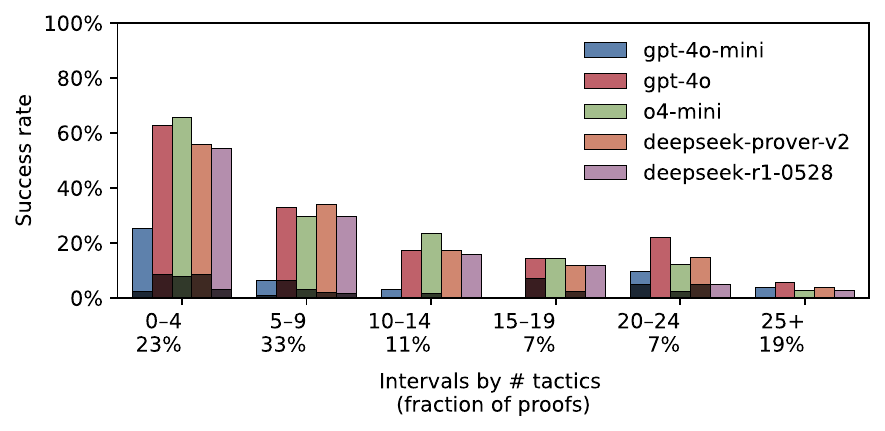}
    \caption{Verdi}\label{fig:verdi-samegraph}
\end{subfigure}
\caption{Success rates (light) vs. identically generated proofs (dark) by tactic count intervals for \hstocoq and Verdi.}
\end{figure}

\Cref{fig:basethy-samegraph,fig:verdi-samegraph} show the proof generation
success rates in each tactic count interval in light colors. These figures show
that, with the exception of GPT-4o-mini, all LLMs have high success rates in
generating proofs of small sizes. These success rates drop as the proof size
increases. However, even when the proof becomes quite large, LLMs can still
succeed in some cases in both projects.

However, one question we need to address to make sure our results are valid is
to check whether LLMs were generating these proofs or whether they have simply
``memorized'' all these proofs, as both projects are open-source projects
available online. For this reason, we further checked if the generated proofs
are identical to the original proofs. We show all the generated identical proofs,
or ``plagiarized'' proofs, in \cref{fig:basethy-samegraph,fig:verdi-samegraph}
using dark colors.

The results show that LLMs indeed generate identical proofs in both projects.
In \hstocoq, these are all small proofs, which have a high likelihood of being
identical ``by coincidence''. On the other hand, some of the larger generated
proofs in Verdi are identical to the original proofs, suggesting that the proof
might have been in these models' knowledge set.

\paragraph{RQ3: Is there a difference when running LLMs on different
  verification projects?} First, the impact of adding dependencies or in-file
context also varies between these two projects. As seen in
\cref{tab:impr-basethy}, the benefits of in-file context diminished in \hstocoq
for proofs involving a larger number of tactics, and, in some cases, even
reduced success rates for certain models. Conversely, simpler proofs with fewer
tactics appeared to benefit from the in-file context.

In contrast, for Verdi, adding in-file context had a remarkably strong effect.
As shown in \cref{tab:impr-verdi}, external dependencies alone were mostly
insufficient for handling longer proofs~(\eg~20+ tactics) with a higher number
of tactics in the original proof. The models were only able to perform better in
the \informed mode, where the in-file context was provided.

\begin{table*}
  \centering
  \caption{Percent gain in success rates from no in-file context (dependencies only) to informed per model and interval (with interval share in \%).}
  \label{tab:improvement}
  \vspace{-0.25cm}
  \begin{subtable}[t]{\linewidth}
    \centering
    \caption{\hstocoq}
    \label{tab:impr-basethy}
    \begin{tabular}{lccccc}
      \toprule
      Model                & 0-4 (56\%)  & 5-9 (21\%)   & 10-14 (12\%) & 15-19 (7\%) & 20+ (5\%)   \\
      \midrule
      GPT-4o               & 16.4 &  0.0  &   9.1  &  -7.7 & 11.1  \\
      GPT-4o-mini          &  8.7 & -2.5  &   0.0  &  -7.7 &  0.0  \\
      o4-mini             & 14.4 & 10.3  & -13.7  &   7.7 & -11.1 \\
      DeepSeek Prover V2   &  0.0 & 18.0  &   9.1  &   0.0 &  0.0  \\
      DeepSeek R1-0528     &  4.8 & 10.3  &   0.0  & -15.4 & 11.1  \\
      \bottomrule
    \end{tabular}
  \end{subtable}%
  \vspace{0.25cm}
  \hfill
  \begin{subtable}[t]{\linewidth}
    \centering
    \caption{Verdi}
    \label{tab:impr-verdi}
    \begin{tabular}{lrrrrrr}
      \toprule
      Model                & 0-4 (23\%)  & 5-9 (33\%)   & 10-14 (11\%)  & 15-19 (7\%) & 20-24 (7\%) & 25+ (19\%)   \\
      \midrule
      GPT-4o-mini          &  7.6 &  4.1  &   3.1  &   0.0 &  9.8  &  3.7  \\
      GPT-4o               &  9.9 & 11.4  &   7.8  &  11.9 & 22.0  &  5.5  \\
      o4-mini              & 10.6 &  6.8  &  12.5  &  11.9 & 12.2  &  2.8  \\
      DeepSeek Prover V2   &  3.8 & 15.1  &  12.5  &   9.5 & 14.6  &  3.7  \\
      DeepSeek R1-0528     &  4.6 &  4.7  &   3.1  &   7.1 &  4.9  &  2.8  \\
      \bottomrule
    \end{tabular}
  \end{subtable}
\end{table*}

Another difference between these two projects is that LLMs did not generate any
proofs identical to original proofs in proofs with a larger tactic count in
\hstocoq.

It is unclear why \hstocoq and Verdi exhibit these differences. However, this
finding suggests that studying one project may not be sufficient for improving
LLMs' effectiveness in other projects.

\paragraph{RQ4: How is the quality of proofs generated by LLMs?}

In this section, we highlight some of the interesting proofs---including both
successful ones and failed ones---generated by LLMs in our case study.

We compare the number of tactics in original proofs and in proofs generated by
LLMs. In both projects, we find that LLMs can generate shorter proofs than
the original ones.

Let's start with an example in \hstocoq. We show an original proof demonstrating
that \cdc{unit} is a monoid that satisfies all the \cdh{Monoid} typeclass laws
in \cref{fig:monoid_unit_original}. The theorem statement itself is not
important. The original proof works by first \cdc{split}ting the theorem into
four subgoals, each representing one monoid property. The proof then unfolds a
number of definitions---a style that is consistent with many other proofs in the
same file. Then, for each subgoal, the proof proceeds by either a case analysis or
an induction.

\begin{figure}[t]
\begin{subfigure}[b]{0.45\textwidth}
\begin{lstlisting}[language=coq]
Instance instance_MonoidLaws_unit :
  MonoidLaws unit.
Proof.
  split;
    unfold op_zlzlzgzg__, Semigroup__unit,
         op_zlzlzgzg____,
         Base.Semigroup__unit_op_zlzlzgzg__;
    unfold mappend, mempty, mconcat,
         Monoid__unit, mappend__,  mconcat__,
         Base.Monoid__unit_mappend,
         Base.Monoid__unit_mempty,
         Base.Monoid__unit_mconcat.
  - intro x. destruct x. auto.
  - intro x. destruct x. auto.
  - intros. auto.
  - intros x. induction x; simpl. auto. auto.
Qed.
\end{lstlisting}
\caption{The original proof showing that \cdc{unit} is a monoid that
  satisfies all the \cdh{Monoid} typeclass laws ~\citep{hs-to-coq, containers}.}\label{fig:monoid_unit_original}
\end{subfigure}
\begin{subfigure}[b]{0.45\textwidth}
\begin{lstlisting}[language=coq]
Proof.
  constructor; intros []; auto.
Qed.
\end{lstlisting}
\caption{A proof for the same theorem generated by OpenAI o4-mini and
  DeepSeek R1-0528. The two models generate the same proof for this
  theorem.}\label{fig:monoid_unit_llm}
\end{subfigure}
\caption{A comparison between the original proof and a proof generated by LLMs
  for theorem \cdc{instance_MonoidLaws_unit} in
  \hstocoq.}\label{fig:monoid-unit-proofs}
\end{figure}

We show a proof generated by OpenAI o4-mini and\\ DeepSeek R1-0528 with \emph{no}
external dependencies or in-file context in \cref{fig:monoid_unit_llm}. The
proof is much simpler: it first uses the \cdc{constructor} tactic, which does
the same thing as \cdc{split} in the original here. Then, LLMs ``realize'' that
all subgoals can be solved using the same sequence of tactics: \cdc{intros []}
to introduce a variable into the context \emph{and} perform a case analysis on
that variable at the same time, then \cdc{auto} for automatically discharging
each goal.

An even smarter proof generated by LLMs can be found in Verdi. We show the
theorem statement in \cref{fig:verdi-ordered}. The theorem describes a relation
between two variables, \cdc{failed} and \cdc{net}, when they are both in a
multi-step transition relation defined by
\cdc{step_ordered_dynamic_failure_star}---the exact definition of this step
relation is not important. The original proof in Verdi is 24 lines of proof
script, involves an induction, and Ltac's \cdc{match} statement.

\begin{figure}[t]
\begin{lstlisting}[language=coq]
Lemma
  ordered_dynamic_state_not_initialized_not_failed : 
forall net failed tr,
  step_ordered_dynamic_failure_star
    step_ordered_dynamic_failure_init
    (failed, net) tr ->
  forall n, ~ In n (odnwNodes net) ->
  ~ In n failed.
(* The following proof is generated
    by OpenAI o4-mini. *)
Proof.
  intros net failed tr Hstar n Hnot Hin.
  apply Hnot.
  eapply ordered_dynamic_failed_then_initialized;
    eauto.
Qed.  
\end{lstlisting}
\caption{A Rocq theorem found in Verdi (in the
  file~\texttt{core/DynamicNetLemmas.v}) and a proof generated by OpenAI
  o4-mini. We omit the original proof found in Verdi because the proof script is
  29 tactics long.}\label{fig:verdi-ordered}
\end{figure}

However, OpenAI o4-mini is able to find a proof that consists of only four basic tactics
in the \informed mode, as shown in \cref{fig:verdi-ordered}. This is because the
contrapositive of this proposition has already been proven as a theorem right
before this theorem (called \cdc{ordered_dynamic_failed_then_initialized}). \\
OpenAI o4-mini ``recognizes'' this connection between the two theorems and
proves this theorem by simply \cdc{apply}ing its contrapositive.

We should point out that the two theorems shown in
\cref{fig:monoid-unit-proofs,fig:verdi-ordered} can also be solved using
classical tools like CoqHammer~\citep{coq-hammer,hammer-semantics}. CoqHammer
can solve the \hstocoq theorem~(\cref{fig:monoid-unit-proofs}) with its own
tactic called \cdc{sfirstorder}. For the Verdi
theorem~(\cref{fig:verdi-ordered}), it performs a proof search using an external
automated theorem prover and also finds that the theorem can be proven with the
help of its contrapositive, similar to the proof generated by LLMs.
Nevertheless, it is impressive that LLMs are able to find these simple proofs
given only one shot without a feedback loop.

The next theorem that LLMs come up with a simpler proof is the most surprising
to us, and the theorem cannot be solved by CoqHammer. We show the theorem and
its original proof in \cref{fig:eqexact_pair_original}. The theorem states that
the pair \cdc{a * b} satisfy the \cdc{EqExact} law~(\cref{fig:base-laws}) if
both \cdc{a} and \cdc{b} satisfies this law. We show the original proof script
in \cref{fig:eqexact_pair_original} to demonstrate the complexity of the
original proof and to compare it with a proof generated by LLMs, but the reader
should not try to read the proof script without Rocq Prover's interactive
environment. The key structure of the proof is to perform two case analyses
indicated by the two uses of the \cdc{destruct} tactic: (1)~if two variables of
type \cdc{a} are equal by \cdc{==}, and (2)~if two variables of type \cdc{b} are
equal by \cdc{==}.

\begin{figure}[t]
\begin{subfigure}[b]{0.45\textwidth}
  \begin{lstlisting}[language=coq]
Instance EqExact_pair {a b}
  `{EqExact a} `{EqExact b} : EqExact (a * b).
Proof.
  split; rewrite /op_zeze__ /op_zsze__
    /Eq_pair___ /op_zeze____ /op_zsze____.
  case =>[??] [??] //=. destruct (_ == _) eqn:?.
  - rewrite andb_true_l. move /Eq_eq in Heqb0.
    destruct (_ == _) eqn:?.
    + constructor. move /Eq_eq in Heqb1.
      subst. reflexivity.
    + constructor. move /Eq_eq in Heqb1.
      intro. apply Heqb1.
      inversion H5; reflexivity.
  - rewrite andb_false_l. constructor.
    move /Eq_eq in Heqb0. intro.
    inversion H5. apply Heqb0. assumption.
Qed.
\end{lstlisting}
\caption{The original proof showing that the pair \cdc{a * b} satisfies the
  \cdc{EqExact} law~(\cref{fig:base-laws}) whenever both \cdc{a} and \cdc{b} do.}\label{fig:eqexact_pair_original}
\end{subfigure}
\begin{subfigure}[b]{0.45\textwidth}
\begin{lstlisting}[language=coq]
Proof.
  split; unfold op_zeze__, op_zsze__,
    Eq_pair___, op_zeze____, op_zsze____
    => - [x1 y1] [x2 y2] /=.
  - case: (Eq_eq x1 x2) => [-> | NEQx];
    case: (Eq_eq y1 y2) => [-> | NEQy];
    constructor; congruence.
Qed.
\end{lstlisting}
\caption{A proof for the same theorem generated by DeepSeek R1-0528.}\label{fig:eqexact_pair_llm}
\end{subfigure}
\caption{A comparison between the original proof and a proof generated by LLMs
  for theorem \cdc{EqExact_pair} in
  \hstocoq.}
\end{figure}

We show a proof generated by DeepSeek R1-0528 with the in-file context but
\emph{without} external dependencies in \cref{fig:eqexact_pair_llm}. The proof
is more concise. A main reason is that, instead of doing a case analysis on
\cdc{==} like in the original proof, this proof performs a case analysis on
\cdc{Eq_eq}, whose type is an inductively defined proposition
\cdc{reflect}~(\cref{fig:base-laws}) that relates \cdc{==} and \cdc{=}.
Intuitively, performing a case analysis on \cdc{reflect} gives us more
information than just \cdc{==}. For example, the original proof uses \cdc{move
  /Eq_eq} to convert between \cdc{==} and \cdc{=} in various places, but such a
step is unnecessary in the generated proof.

Performing a case analysis on an inductively defined pro\-position like
\cdc{reflect} to ``extract more information'' is a classical technique in
mechanized verification discussed in Rocq Prover
textbooks~\citep[Chapter~``Inductively Defined Propositions'']{sf1}. However,
\cdc{Eq_eq} is a function defined in \texttt{hs\--to-coq}---to be more precise,
in the same file as \cdc{EqExact_pair}\\---and a case analysis on \cdc{Eq_eq} was used only
once in the same file before this theorem, but DeepSeek R1-0528 is still able to
generate a proof like this in one shot without a feedback loop.

On the negative side, we find that LLMs can stutter in generating some proofs.
We show a theorem in \hstocoq and its original proof in
\cref{fig:simpl_list_cons_original}. The proof is simple and straightforward.
However, GPT-4o-mini generates a failed proof in the \informed mode shown in
\cref{fig:llm-stutter}. The proof fails at the first \cdc{unfold}. However, an odder
phenomenon is that, after the first \cdc{rewrite /op_zeze__}, the proof
just keeps repeating the same tactic, seemingly indefinitely.

\begin{figure}[t]
\begin{subfigure}[b]{0.45\textwidth}
\begin{lstlisting}[language=coq]
Lemma simpl_list_cons_eq a `{Eq_ a}
  (x y :a) xs ys :
  (cons x xs) == (cons y ys)
  = (x == y) && (xs == ys).
Proof.
    unfold op_zeze__, op_zeze____, Eq_list.
    simpl. auto.
 Qed.
\end{lstlisting}
\caption{The theorem \cdc{simpl_list_cons_eq} theorem in \hstocoq and its original proof.}\label{fig:simpl_list_cons_original}
\end{subfigure}
\begin{subfigure}[b]{0.45\textwidth}
\begin{lstlisting}[language=coq]
(* The following proof is generated by
  GPT 4o-mini. *)
Proof.
  unfold Eq___list, op_zeze__, op_zeze____, eqlist.
  rewrite /eqlist.
  destruct xs, ys; simpl.
  - rewrite /op_zeze__.
    rewrite /op_zeze__.
    rewrite /op_zeze__.
    (* The proof goes on by
      repeating [rewrite]s. *)
\end{lstlisting}
\caption{A failed proof generated by GPT 4o-mini that repeats the same
  tactic.}\label{fig:llm-stutter}
\end{subfigure}
\caption{A comparison between the original proof and a failed proof generated by
  LLMs for theorem \cdc{simpl_list_cons_eq} in \hstocoq.}
\end{figure}

This phenomenon is not common but has appeared in both \hstocoq and Verdi across
different theorems and LLMs when invoking various tactics. A low temperature
(0.1) is a possible factor, but we did not run further experiments to test this
hypothesis.

\paragraph{Limitations}

In terms of design, our experimental setup focused exclusively on single-pass
proof generation, without incorporating a feedback loop. Although we recorded
errors produced by \lstinline|sertop| for each LLM and setting for future work,
we did not analyze the model's ability to respond to errors. Moreover, when
analyzing dependencies, we limited our scope to files within the same Rocq
project. While the models were given the import statements in each setting, the
external libraries were not indexed. Consequently, we assumed that LLMs would
have implicit knowledge of widely used libraries such as ssreflect, or
StructTact, InfSeqExt, and Cheerios (for Verdi), but did not verify this
directly.

A further limitation lies in our dataset, which, while containing a substantial
number of theorems, covers only two Rocq projects. Rocq projects may naturally
vary in their structure and organization, which may heavily impact the results
for settings with one of the dependencies or in-file contexts.

Finally, our experiments were conducted using Rocq version 8.10.2 and 8.11.0,
which are both relatively old. While this choice was necessary to ensure
compatibility with the codebases we studied, it may impact the relevance of
results for newer versions of Rocq if the LLMs we used were trained on more
recent versions of the language.

\section{Related Work}\label{sec:related-work}

\paragraph{Benchmarks for proofs}
CoqGym is a pioneer in providing an extensive Rocq benchmark for machine
learning models~\citep{coq-gym}, containing 71K proofs from 123
real-life projects. It has been used by various studies on proof automation, such
as \citet{palm,tactok,diversity-driven}. These works are also an inspiration for
the case studies presented in this paper. However, one issue with CoqGym is that it
relies on older versions of Rocq Prover. For this reason, more recent tools like
the \texttt{CoqPilot} benchmarking framework choose to build their own
datasets~\citep{coq-pilot}.





Outside Rocq Prover, there are many benchmarks for other proof assistants or
formal-method tools, such as Dafny\-Bench~\citep{dafny-bench},
LeanDojo~\citep{lean-dojo}, miniCodeProps~\citep{minicodeprops},
FVAPPS \citep{fvapps}, VerifyThisBench~\citep{verify-this-bench},
Verina~\citep{verina}, \etc\@

\paragraph{Proof automation}

Proof automation has always been a goal in research on proof assistants. Most of
these works rely on automated theorem provers~(ATPs) like SAT/SMT solvers. For
example, SMTCoq~\citep{smt-coq} uses SAT/SMT solvers to prove theorems and then
reconstructs Rocq proofs from them. CoqHammer~\citep{coq-hammer,hammer-semantics}
defines a set of automation tactics for dependent type theory, uses external
ATPs to find a proof, and then constructs a proof using its automation
tactics by taking \emph{hints} from proofs found by ATPs. In this way, CoqHammer
is able to construct Rocq proofs that use intuitionistic logic with the help of
ATPs that work on classical logic.

Other proof automation tools like Tactician~\citep{tactician} use machine
learning (but not LLMs) instead. It provides suggestions for the next tactic
based on ``previously written tactics''. CoqGym, the benchmark for Rocq proofs,
also includes a tool called ASTactic, which is trained on CoqGym and uses
deep learning to generate proofs automatically~\citep{coq-gym}. Some more recent
works in this area include Proverbot9001~\citep{proverbot9001},
Passport~\citep{passport}, QEDCartographer~\citep{qed-cartographer}, \etc\@

\paragraph{LLMs and proof assistants}
There have been a few recent works that investigate the capabilities of LLMs in
generating proofs for proof assistants. We have already discussed \citet{palm}'s
study on Verdi in \cref{sec:verdi}. \Citet{fscq-llm} studied FSCQ, a verified
file system~\citep{fscq}. They conjectured that one reason LLMs fail to generate
proofs is that LLMs struggle to find relevant lemmas when too many lemmas are
given in a prompt~\citep[Section 4.3]{fscq-llm}.

There have also been many works that leverage the power of LLMs to build
proof-automation tools. For example, Baldur uses fine-tuned LLMs to generate
whole proofs for Isabelle/HOL~\citep{baldur}. Their evaluation of Baldur on the
PISA dataset~\citep{pisa} further shows that LLMs outperform small-model-driven
search-based methods. \emph{PALM} builds on its observation on
Verdi~(\cref{sec:verdi}) and uses a generate-then-repair approach that combines
LLMs and symbolic methods (\eg~CoqHammer~\citep{coq-hammer,hammer-semantics}) to
generate Rocq proofs~\citep{palm}. Draft, Sketch, and Prove~(DSP) uses LLMs to
generate a sketch of a formal proof and then uses ATPs to fill in the missing
details in the sketch~\citep{dsp}. Some other works in this area include
\citet{hybrid-prover,cobblestone,selene,fvel,rango},~\etc\@

\paragraph{Premise selection for proof generation}
Premise selection refers to the process of selecting relevant \emph{premises},
such as definitions and lemmas~\citep{deepmath}. This is a common process used
by many proof-generation works. For example, \emph{PALM} uses Term
Frequency-Inverse Document Frequency~(TF-IDF)\\~\citep{tf-idf} and $k$ nearest
neighbors~(KNN)~\citep{knn} to select relevant premises. \texttt{CoqPilot}
selects premises based on ``metrics such as distance from the generation target
or similarity with other theorem statements''~\citep{coq-pilot}.

Our work takes a much simpler approach by directly including dependencies and
in-file context in the prompt. Prior works like Baldur did a similar thing, but
they only included in-file context~\citep[Section 2.3]{baldur}.



\paragraph{LLMs and math}

LLMs have been studied extensively in the context of mathematics. Earlier
research focuses on benchmarking LLMs with simple math reasoning
tasks~\cite{GSM8k,miniF2F,ProofNet}. Recently, Olympiad-level math theorem
proving has been successfully tackled by LLMs~\cite{AlphaGeometry,
  AlphaProof,OlympiadIneq}. There has also been rapid progress in
auto-formalizing
mathematics~\cite{autoformalize-neurips22,autoformalize-icml24,autoformalize-neurips24}.


\section{Conclusion}\label{sec:conclusion}

In this paper, we conduct a case study based on two real-world Rocq projects:
the \hstocoq project and Verdi. Our case study shows that LLMs can be effective
in generating whole proofs for program correctness theorems. More specifically,
we show that external dependencies and in-file context can significantly help
with proof generation. We also find that LLMs perform well on small proofs.
While its effectiveness degrades when the proof size increases, there is still a
decent chance for it to generate whole proofs. However, our study also shows
that the effectiveness characteristics of LLMs differ in different verification
projects, which suggests that studying one project may not be sufficient for
improving LLMs' effectiveness in other projects. Finally, we find that LLMs can
generate concise and smart proof scripts, can apply classical techniques to new
definitions, but can also produce meaningless stuttering proofs.

We believe that using LLMs for verification with proof assistants is a
promising direction that deserves more attention. Program verification is
suitable for tools like LLMs that are unpredictable and can
hallucinate~\citep{hallucination,more-llm-proofs}. First, proofs are \emph{not}
computational. A generated \emph{inefficient} proof has little to no impact
compared with a generated inefficient program. Second, the proof-checking
mechanisms in proof assistants~(\eg~\cdc{Qed} of Rocq Prover) can safeguard
generated proofs to make sure that they are correct.

Verification with proof assistants can be potentially much more useful in
software engineering if proof automation can be significantly improved. Indeed,
researchers have argued that one major reason that formal methods are rarely
used in software development today is their social aspect~\citep{social-fm}. It
will greatly improve the usability of formal methods (and hence the reliability of
software) if LLMs can help with proof automation.

\section*{Data-Availability Statement}
All code, scripts, logs, and the container image needed to reproduce our results are available on GitHub (\url{https://github.com/bbayazit16/lmpl-2025-artifact}) and are archived on Zenodo at \href{https://doi.org/10.5281/zenodo.16939067}{10.5281/zenodo.16939067}~\citep{bayazit}.

\begin{acks}

  We thank all the anonymous reviewers of LMPL 2025 for their thoughtful and
  constructive comments on this paper and their suggestions for potential future
  directions for this work. We thank Yiming Lin for his feedback on a draft of
  this paper.

  This work was partially supported by the University of Toronto Department of
  Computer Science Research Award. This work has also partially benefited from
  the Microsoft Accelerate Foundation Models Research (AFMR) grant program.

\end{acks}

\balance
\bibliography{ref}

\end{document}